\documentclass[preprintnumbers,floatfix,12pt,superscriptaddress,nofootinbib]{revtex4}

\usepackage{amsfonts}
\usepackage{amssymb,epsfig}
\usepackage{latexsym}
\usepackage{amsmath}
\usepackage{graphicx}

\begin{document}

\title{Cosmological implications of interacting polytropic gas dark energy model in non-flat universe}
\author{M. Malekjani $^{1,2}$\footnote{%
Email: \text{malekjani@basu.ac.ir}},A. Khodam-Mohammadi $^{1}$\footnote{%
Email: \text{khodam@basu.ac.ir}} and M. Taji}
\affiliation{1-Department of Physics, Faculty of Science, Bu-Ali
Sina University, Hamedan 65178, Iran\\
2-Research Institute for Astronomy and Astrophysics of Maragha
(RIAAM), Maragha, Iran}

\begin{abstract}
\vspace*{1.5cm} \centerline{\bf Abstract} \vspace*{1cm} The
polytropic gas model is investigated as an interacting dark energy
scenario. The cosmological implications of the model including the
evolution of EoS parameter $w_{\Lambda}$, energy density
$\Omega_{\Lambda}$ and deceleration parameter $q$ are investigated.
We show that, depending on the parameter of model, the interacting
polytropic gas can behave as a quintessence or phantom dark energy.
In this model, the phantom divide is crossed from below to up. The
evolution of $q$ in the context of polytropic gas dark energy model
represents the decelerated phase at the early time and accelerated
phase later. The singularity of this model is also discussed.
Eventually, we establish the correspondence between interacting
polytropic gas model with tachyon, K-essence and dilaton scalar
fields. The potential and the dynamics of these scalar field models
are reconstructed according to the evolution of interacting
polytropic gas.
\end{abstract}

\maketitle

\newpage
\section{Introduction}
Recent cosmological observations obtained by SNe Ia \cite{c1}, WMAP
\cite{c2}, SDSS \cite{c3} and X-ray \cite{c4} experiments reveal
that our universe expands under an accelerated expansion. In the
framework of standard Freidmann-Robertson-Walker (FRW) cosmology, a
missing energy component with negative pressure dubbed dark energy
(DE) is responsible for this expansion. The nature of DE is still
unknown and scientists believe that the problem of DE is a major
puzzle of modern cosmology. Up to now, many theoretical models have
been investigated to interpret the behavior of DE. The
time-independent cosmological constant, $\Lambda$, with EoS
parameter $w=-1$ is the earliest and simplest candidate of DE. The
cosmological constant suffers from two well known difficulties
namely "fine-tuning" and "cosmic coincidence" problems. The
alternative candidates for DE problem are the dynamical dark energy
scenario with time varying EoS parameter ,$w$. According to some
analysis on the SNe Ia observational data, it has been shown that
the time-varying DE models give a better fit compare with a
cosmological constant \cite{c44}. There are two different categories
for dynamical DE scenario: (\emph{i}) The scalar fields including
quintessence \cite{c5}, phantom \cite{c6}, quintom \cite{c7},
K-essence \cite{c8}, tachyon \cite{c9}, dilaton \cite{c10} and so
forth. (\emph{ii}) The interacting DE models including Chaplygin gas
models \cite{c100, setare22}, braneworld models \cite{101},
holographic \cite{c11} and agegraphic \cite{c12} models. The
holographic DE model is constructed in the light of holographic
principle of quantum gravity \cite{c13} and the agegraphic model is
constructed based on the uncertainty relation of quantum mechanics
together with the gravitational effect in general relativity
\cite{c14}. The interaction between DE and dark matter is supported
by recent observations prepared by the Abell Cluster A586
\cite{c15}. However the strength of this interaction is not clearly
identified \cite{c16}. Also, recent astronomical data supported that
our universe is not a perfectly flat and has a small positive
curvature
\cite{c166}.\\
The polytropic gas model has an important role in stellar
astrophysics. It can explain the equation of state of degenerate
electrons and degenerate neutrons in white dwarfs and neutron stars,
respectively \cite{c19}. This model can also be useful when the
pressure and density are adiabatically related to each other in main
sequence stars \cite{c19}. Here we consider the interpretation of
 dark energy scenario with the EoS parameter of polytropic gas.
U. Mukhopadhyay and S. Ray used some dynamical $\Lambda$ model with
polytropic equation of state in dark energy scenario \cite{ray05}.
Recently, by using the polytropic gas model, the interaction between
DE and dark matter is investigated \cite{c17}. Karami, et al.
obtained the phantom behavior of interacting polytropic gas model
\cite{c17}. Also, karami, et al. reconstructed the $f(T)$-gravity
from the polytropic gas DE model \cite{c18}. They also studied the
correspondence between the interacting new agegraphic dark energy
model with polytropic gas model in non-flat FRW universe and
reconstructed the potential and the dynamics for the scalar field of
the polytropic model to describe the accelerated expansion of the
universe \cite{karam20}. The above statements motivate us to
consider more cosmological implications of this model in dark energy
scenario. One of the interesting features of this model that we
discuss is that in the polytropic gas dark energy scenario the
phantom regime can be achieved even in the absence of interaction
between dark energy and dark matter. This makes it distinguishable
from many other DE model whose $W_{\Lambda}$ can not crosses the
phantom regime without the interaction between DE and dark matter.
We consider the interacting polytropic gas as a phenomenological DE
model. In the phenomenological models of DE the pressure $p$ is
given as a function of energy density $\rho$, i.e.,
$p=-\rho-f(\rho)$ \cite{c177}. Considering $f(\rho)=0$, the EoS
 parameter of phenomenological models cross $w=-1$, i.e., the EoS of
 cosmological constant. Nojiri, et al. investigated four types
 singularities for some illustrative examples of phenomenological
 models \cite{c177}. The polytropic gas model has a type III.
 singularity in which the singularity takes place at a
 characteristic scale factor $a_s$.\\
Here, we obtain the deceleration parameter $q$ to explain the
decelerated and accelerated expansion phases of the universe
dominated by polytropic gas dark energy fluid. The behavior of
interacting polytropic gas in the quintessence regime is also
calculated. We study the correspondence between the tachyon,
K-essence and dilaton fields with the interacting polytropic gas
dark energy and reconstruct the potential and the dynamics of these
scalar fields according the evolutionary form of interacting
polytropic gas model.
\section{Polytropic gas DE model}
 The equation of state (EoS) of polytropic gas is
given by
\begin{equation}\label{poly}
p_{\Lambda}=K\rho_{\Lambda}^{1+\frac{1}{n}},
\end{equation}
where $K$ and $n$ are the polytropic constant and polytropic index,
respectively \cite{c19}.\\Assuming a non-flat
Friedmann-Robertson-Walker (FRW) universe containing  DE and CDM
components, the corresponding Friedmann equation is as follows
\begin{equation}\label{frid1}
H^{2}+\frac{k}{a^{2}}=\frac{1}{3M_{p}^{2}}(\rho _{m}+\rho
_{\Lambda})
\end{equation}%
where $H$ is the Hubble parameter, $M_p$ is the reduced Plank mass
and $k=1,0,-1$ is a curvature parameter corresponding to a closed,
flat and open universe, respectively. $\rho_m$ and $\rho_{\Lambda}$
are the energy density of CDM and DE, respectively. Recent
observations support a closed universe with a tiny positive small
curvature $\Omega _{k}=\simeq
0.02$ \cite{c20}.\\
The dimensionless energy densities are defined as
\begin{equation}\label{denergy}
\Omega_{m}=\frac{\rho_m}{\rho_c}=\frac{\rho_m}{3M_p^2H^2}, ~~~\\
\Omega_{\Lambda}=\frac{\rho_{\Lambda}}{\rho_c}=\frac{\rho_{\Lambda}}{3M_p^2H^2}~~\\
\Omega_k=\frac{k}{a^2H^2}
\end{equation}
Therefore the Friedmann equation (\ref{frid1}) can be written as
\begin{equation}
\Omega _{m}+\Omega _{\Lambda}=1+\Omega _{k}.  \label{Freq2}
\end{equation}%
 Considering a universe dominated by interacting polytropic gas DE and
CDM, the total energy density, $\rho=\rho_m+\rho_{\Lambda}$,
satisfies a conservation equation
\begin{equation}
\dot{\rho}+3H(\rho+p)=0
\end{equation}
However, by considering the interaction between DE and dark matter,
the energy density of DE and dark matter does not conserve
separately and in this case the conservation equations are given by
\begin{eqnarray}
\dot{\rho _{m}}+3H\rho _{m}=Q, \label{contm}\\
\dot{\rho _{\Lambda}}+3H(\rho_{\Lambda}+p_{\Lambda})=-Q,
\label{contd}
\end{eqnarray}%
where  $Q$ indicates the interaction between DE and CDM. Three forms
of $Q$ which have been extensively used in the literatures are
\cite{c21}
\begin{equation}\label{Q-term}
Q=3 \alpha H\rho_{\Lambda}, ~~~ 3\beta H\rho_m, ~~~ 3\gamma
H(\rho_{\Lambda}+\rho_m),
\end{equation}
where  $\alpha$, $\beta$ and $\gamma$ are the dimensionless
constants. The Hubble parameter $H$ in the $Q$-terms is considered
for mathematical simplicity. Indeed, the interaction forms in
Eq.(\ref{Q-term}) are given by hand, since the $Q$ in
Eqs.(\ref{contm}, \ref{contd}) should be as a function of $H$
multiplied with energy density. Similar to the standard $\Lambda$CDM
model, in which the vacuum fluctuations can decay into matter, here
the interaction parameter $Q$ indicates the decay rate of the
polytropic gas into CDM component. Recently, the the interaction
between DE and dark matter is presented in \cite{c233}. For
mathematical simplicity, we consider the first form of interaction
parameter $Q$.\\ 
Using Eq.(\ref{poly}), the integration of
 continuity equation for interacting dark energy component, i.e. Eq.(\ref{contd}), obtains
 \begin{equation}\label{rho1}
 \rho_{\Lambda}=\left(\frac{1}{Ba^{\frac{3(1+\alpha)}{n}}-\widetilde{K}}\right)^n,
 \end{equation}
where $B$ is the integration constant,
$\widetilde{K}=\frac{K}{1+\alpha}$ and $a$ is the scale factor. Note
that to have a positive energy density for an arbitrary number of
$n$, it is required $Ba^{3(1+\alpha)/n}>\widetilde{K}$. It is
worthwhile to note that the phantom behavior of interacting
polytropic gas has been also studied in  \cite{c17}. In the case of
$Ba^{3(1+\alpha)/n}=\widetilde{K}$, we have $\rho\rightarrow \infty$
and therefore the polytropic gas has a finite-time singularity at
$a_c=(\widetilde{K}/B)^{n/3(1+\alpha)}$. This type of singularity,
in which at a characteristic scale factor $a_s$, the
 energy density $\rho\rightarrow\infty$ and the pressure density
 $|p|\rightarrow\infty$, is indicated by type III singularity
\cite{c177}.\\
Substituting $Q=3\alpha H\rho_{\Lambda}$ in (\ref{contd}), we have
\begin{equation}\label{contd2}
\dot{\rho _{\Lambda}}+3H(1+\alpha+w_{\Lambda})\rho_{\Lambda}=0,
\end{equation}
Taking the derivative of Eq.(\ref{rho1}) with respect to time, one
can obtain
\begin{equation}\label{dotrho}
\dot{\rho_{\Lambda}}=-3BH(1+\alpha)a^{\frac{3(1+\alpha)}{n}}\rho_{\Lambda}^{1+\frac{1}{n}}
\end{equation}
Substituting Eq.(\ref{dotrho}) in (\ref{contd2}) and using
Eq.(\ref{rho1}) , we can obtain the  EoS parameter of interacting
polytropic gas as
\begin{equation}\label{eos1}
 w_{\Lambda}=-1-\frac{a^{\frac{3(1+\alpha)}{n}}}{c-a^{\frac{3(1+\alpha)}{n}}}-\alpha
\end{equation}
 where $c=\widetilde{K}/B$. By defining the effective EoS parameter as
  $w_{\Lambda}^{eff}=w_{\Lambda}+\alpha=-1-a^{\frac{3(1+\alpha)}{n}}/(c-a^{\frac{3(1+\alpha)}{n}})$
 , we see that the interacting polytropic gas model behaves as a
phantom
 model, i.e.
 $w_{\Lambda}^{eff}<-1$, when $c>a^{3(1+\alpha)/n}$. The phantom behavior of polytropic gas is similar to
 generalized chaplygin gas model, where it has been shown that the generalized chaplygin gas
 with negative value of model parameter can behave as a phantom dark energy \cite{setare22}.
    Note that in the case of phantom polytropic gas, from Eq.(\ref{rho1})
  we see that only even numbers of $n$ should be chosen to have a positive energy density.
The interesting feature of polytropic gas model is that it can
obtain the phantom regime even in the absence of interation. For
this aim, it is enough to insert $\alpha=0$ in Eq.(\ref{eos1}) and
see that for $c>a^{\frac{3(1+\alpha)}{n}}$ the phantom regime,
$w_{\Lambda}<-1$, can be achieved. This makes it distinguishable
from many other dark energy models whose $w_{\Lambda}$ cannot cross
the phantom regime without interaction term. The other interesting
aspect of the polytropic gas is that the interacting polytropic gas
dark energy crosses the phantom divide from $w_{\Lambda}<-1$ to
$w_{\Lambda}>-1$ (see Fig.(1), left panels). This behavior of
polytropic gas is similar to interacting agegraphic dark energy
model in which the phantom divide is crossed from below to up (see
Figs.(2,3) of \cite{cai44}). The similarity of the interacting
agegraphic dark energy and polytropic gas is that both models cross
the phantom divide from below to up.\\
 The interacting polytropic gas
behaves as a quintessence model, i.e. $-1<w_{\Lambda}^{eff}<-1/3$,
when $-\infty<c\leq-a^{3(1+\alpha)/n}/2$. The condition
$-a^{3(1+\alpha)/n}/2<c<a^{3(1+\alpha)/n}$ leads to
$w_{\Lambda}^{eff}>-1/3$ and consequently the accelerated expansion,
in this case, can not be achieved. At $c=a^{3(1+\alpha)/n}$, the
interacting polytropic gas has a singularity. Hence, depending on
the parameter $c$, the polytropic gas can behaves as a phantom or
quintessence models of DE. Also it is worth to mention that the
polytropic gas model behaves as a cosmological constant,
i.e.,$w_{\Lambda}^{eff}\rightarrow-1$, at the early time (i.e.
$a\rightarrow0$) whereas the universe is dominated
by pressureless dark matter.\\
In Fig. (1), the evolution of $w_{\Lambda}$ as a function of scale
factor is plotted for different values of the parameters $c$ and
$n$.The other interesting aspect of the polytropic gas is that the
interacting polytropic gas dark energy crosses the phantom divide
from $w_{\Lambda}<-1$ to $w_{\Lambda}>-1$ (see Fig.(1), left
panels). This behavior of polytropic gas is similar to interacting
agegraphic dark energy model in which the phantom divide is crossed
from below to up (see Figs.(2,3) of \cite{cai44}). The similarity of
the interacting agegraphic dark energy and polytropic gas is that
both models cross the phantom divide from below to up. In upper
panels we fix the polytropic index as $n=2$ and in lower panels the
parameter $c$ is fixed. In upper left panel the negative values of
$c$ are selected to obtain the transition from phantom to
quintessence regime. In upper right panel, the positive values of
$c$ are selected. In this case the interacting polytropic gas
behaves as a phantom like field. Same as left panel, we fix the
polytropic index $n=2$. Here, one can easily find the phantom
behavior of polytropic gas model. It is worth noting that the
phantom regime of polytropic gas model is restricted with a
characteristic scale factor $a_s=c^{n/3(1+\alpha)}$, where we
encounter with a singularity at this epoch. In lower panels of
Fig.(1), the dependency of the evolution of $w_{\Lambda}$ on the
polytropic index parameter $n$ is studied. In lower left panel, by
fixing $c=-1$, we studied this dependency for polytropic gas model.
It is easy to see that the larger value of $n$ gets the larger
$w_{\Lambda}$ at $a<1$ and smaller $w_{\Lambda}$ at $a>1$. In lower
right panel, by fixing $c=2$, the dependency of $w_{\Lambda}$ on the
parameter $n$ is investigated for phantom polytropic gas model.
Unlike to lower left panel, the larger value of $n$ gets the smaller
$w_{\Lambda}$ at $a<1$ and larger $w_{\Lambda}$ at $a>1$.\\

In order to obtain the evolution of dimensionless energy density,
$\Omega_{\Lambda}$, let us start with Eqs.(\ref{rho1}) and
(\ref{denergy}) and obtain the density parameter of interacting
polytropic gas as
\begin{equation}\label{denergy2}
\Omega_{\Lambda}=\frac{(Ba^{\frac{3(1+\alpha)}{n}}-\widetilde{K})^{-n}}{3M_p^2H^2}
\end{equation}
Taking the derivative of Eq.(\ref{denergy2}) with respect to time
and using $\Omega^{\prime}=\dot{\Omega}/H$, we can obtain
\begin{equation}\label{motion}
\Omega_{\Lambda}^{\prime}=-\Omega_{\Lambda}\Big(\frac{3(1+\alpha)a^{\frac{3(1+\alpha)}{n}}}{a^{\frac{3(1+\alpha)}{n}}-c}+2\frac{\dot{H}}{H^2}\Big)
\end{equation}
where prime denotes the derivative with respect to $x=\ln{a}$.
Taking the derivative of Friedmann equation (\ref{frid1}) with
respect to time and using Eqs.(\ref{rho1}), (\ref{Freq2}),
(\ref{contm}), (\ref{denergy2}) and $Q=3\alpha H\rho_{\Lambda}$, one
can find that
\begin{equation}\label{doth}
\frac{\dot{H}}{H^2}=-\frac{3}{2}\Big[\Omega_{\Lambda}\frac{c(1+\alpha)}{a^\frac{3(1+\alpha)}{n}-c}+1+\frac{\Omega_k}{3}\Big]
\end{equation}
Substituting this relation into Eq.(\ref{motion}), we obtain the
evolutionary equation for energy density parameter of interacting
polytropic gas as:
\begin{equation}\label{omega_evol}
\Omega_{\Lambda}^{\prime}=-3\Omega_{\Lambda}\Big[\frac{c}{a^{\frac{3(1+\alpha)}{n}}-c}(1-\Omega_{\Lambda})+\alpha
\frac{a^{\frac{3(1+\alpha)}{n}}-c\Omega_{\Lambda}}{a^{\frac{3(1+\alpha)}{n}}-c}-\frac{\Omega_k}{3}\Big],
\end{equation}
where $\Omega_k$ is given by
\begin{equation}
\Omega_k=a\gamma\frac{1-\Omega_{\Lambda}}{1-a\gamma}
\end{equation}
and $\gamma=\Omega_{k0}/\Omega_{m0}$.\\
 In Fig.(2), by solving the differential
equation (\ref{omega_evol}), we show the evolution of
$\Omega_{\Lambda}$ for different model parameters $c$ and $n$ as
well as different interaction parameter $\alpha$. Here we assume
only the positive values of $c$, i.e., the phantom polytropic gas
model. The numerical values of density parameters at the present
time are taken as: $\Omega_{\Lambda0}=0.7$, $\Omega_{m0}=0.3$ and
$\Omega_{k0}=0.02$. In upper panels, we consider the non-interacting
polytropic gas and in lower panel the interaction term is included.
Here, we see that $\Omega_{\Lambda}\rightarrow 0$ at the early time
and tends to $1$ at the late time. Hence the polytropic gas model
can describe the matter-dominated universe in the far past. Also, at
the late time, we encounter with dark energy dominated universe
($\Omega_{\Lambda}\rightarrow 1$). In upper left panel, by fixing
the parameter $n$, the polytropic gas starts to be effective earlier
and $\Omega_{\Lambda}$ tends to a lower value at the late time when
$c$ is larger. On the other hand, in upper right panel, we see that
for fixed parameter $c$, the polytropic gas starts to be effective
earlier and also $\Omega_{\Lambda}$ tends to a higher value at the
late time when $n$ is smaller. In lower panel, the effect of
interaction parameter $\alpha$ on the evolution of
$\Omega_{\Lambda}$ is studied. Here one can see that the polytropic
gas starts to be effective earlier, by increasing the interaction
parameter $\alpha$. Also, at $a>1$, the parameter $\Omega_{\Lambda}$
is smaller for
larger values of $\alpha$.\\
 For completeness, we derive the deceleration parameter
\begin{equation}\label{qdece}
q=-\frac{\ddot{a}}{aH^2}=-1-\frac{\dot{H}}{H^2}
\end{equation}
for polytropic gas model. Substituting Eq.(\ref{doth}) in
(\ref{qdece}) we get
\begin{equation}\label{qdece2}
q=-1+\frac{3}{2}\Big[\Omega_{\Lambda}\frac{c(1+\alpha)}{a^\frac{3(1+\alpha)}{n}-c}+1+\frac{\Omega_k}{3}\Big]
\end{equation}
It is worth noting that in the limiting case of matter-dominated
phase and considering flat universe in the absence of interaction
term, Eq.(\ref{qdece2}) is reduced to $q=1/2$ which represents the
decelerated expansion ($q>0$) of the universe.\\
In Fig.(3), we show the evolution of $q$ as a function of $a$ for
different model parameters $c$ and $n$ as well as different
interaction parameter $\alpha$. Here we discuss the evolution of $q$
for phantom polytropic gas model, by assuming positive $c$. Upper
panels is plotted in the absence of interaction between dark energy
and dark matter
 and the lower panel is plotted in the presence of interaction term.
 The parameter $q$ converges to $1/2$ at the early time, whereas the universe
is dominated by pressureless dark matter. In upper left panel, by
fixing $n$, the accelerated expansion is achieved earlier by
increasing $c$. Also, in the upper right panel, we see that by
increasing $n$, $q$ becomes larger at the deceleration phase and
gets smaller at the acceleration phase. It is worth noting that,
although, both the model parameters $n$ and $c$ impact the evolution
of deceleration parameter $q$, but the change of sign from $q>0$ to
$q<0$ depends on the parameter $c$ of the model.
\section{Correspondence between polytropic gas DE model and scalar fields}
In the present section we establish a correspondence between the
interacting polytropic gas model with the tachyon, K-essence and
dilaton scalar field models. The importance of this correspondence
is that the scalar field models are an effective description of an
underlying theory of dark energy and therefore it is worthwhile to
reconstruct the potential and the dynamics of scalar fields
according the evolutionary form of polytropic gas model. For this
aim, first we compare the energy density of polytropic gas model
(i.e. Eq.\ref{rho1}) with the energy density of corresponding scalar
field model. Then, we equate the equations of state of scalar field
models with the EoS parameter of polytropic gas (i.e.
Eq.\ref{eos1}).
\subsection{Polytropic gas tachyon model}
It is believed that the tachyon can be assumed as a source of DE
\cite{c24}. The tachyon is an unstable field which can be used in
string theory through its role in the Dirac-Born-Infeld (DBI) action
to describe the D-bran action \cite{c25}. The effective Lagrangian
for the tachyon field is given by
\[
\mathcal{L}=-V(\phi )\sqrt{1-g^{\mu \nu }\partial _{\mu }\phi
\partial _{\nu }\phi },
\]%
where $V(\phi )$ is the potential of tachyon. The energy density and
pressure of tachyon field are  \cite{c25}
\begin{equation}
\rho _{\phi }=\frac{V(\phi )}{\sqrt{1-\dot{\phi}^{2}}},
\label{tach1}
\end{equation}%
\begin{equation}
p_{\phi }=-V(\phi )\sqrt{1-\dot{\phi}^{2}}.
\end{equation}%
The EoS parameter of tachyon can be obtained as
\begin{equation}
w_{\phi }=\frac{p_{\phi }}{\rho _{\phi }}=\dot{\phi}^{2}-1.
\label{eos_tach}
\end{equation}%
In order to have a real energy density for tachyon field, it is
required that $-1<\dot{\phi}<1$. Consequently, from
Eq.(\ref{eos_tach}), the EoS parameter of tachyon is constrained to
$-1<w_{\phi}<0$. Hence, the tachyon field can interpret the
accelerates expansion of universe, but it can not enter the phantom
regime, i.e. $w_{\Lambda}<-1$. In order to reconstruct the potential
and the dynamics of tachyon according to evolution of interacting
polytropic gas model, we should equate Eqs.(\ref{eos1}) and
(\ref{eos_tach}) and also Eq.(\ref{rho1}) with Eq.(\ref{tach1}) as
follows

\begin{equation}
w_{\Lambda}=-1-\frac{a^{\frac{3(1+\alpha)}{n}}}{c-a^{\frac{3(1+\alpha)}{n}}}-\alpha=\dot{\phi}^{2}-1.
\end{equation}

\begin{equation}
\rho_{\Lambda}=\left(\frac{1}{Ba^{\frac{3(1+\alpha)}{n}}-\widetilde{K}}\right)^n=\frac{V(\phi
)}{\sqrt{1-\dot{\phi}^{2}}}
\end{equation}
Hence we get the following expressions for dynamics and potential of
tachyon field
\begin{equation}\label{dotphi1}
\dot{\phi}^{2}=-\frac{a^{\frac{3(1+\alpha)}{n}}}{c-a^{\frac{3(1+\alpha)}{n}}}-\alpha
\end{equation}

\begin{equation}
V(\phi)=\sqrt{1+\frac{a^{\frac{3(1+\alpha)}{n}}}{c-a^{\frac{3(1+\alpha)}{n}}}+\alpha}
\left(\frac{1}{Ba^{\frac{3(1+\alpha)}{n}}-\widetilde{K}}\right)^n
\end{equation}
For $c>a^{3(1+\alpha)/n}$, from Eq.(\ref{dotphi1}), we obtain
$\dot{\phi}^{2}<0$ which represents the phantom behavior of tachyon
field. It is worth noting that the reconstructed tachyon field
according to the interacting polytropic gas can cross the phantom
divide. By definition $\phi=i\psi$ and changing the time derivative
to the derivative with respect to logarithmic scale factor, i.e.
$d/dt=H d/dx$, the scalar field $\psi$ can be integrated from
Eq.(\ref{dotphi1}) as follows
\begin{equation}
\psi(x)-\psi(0)=\int_0^x{\frac{1}{H}\sqrt{-\frac{a^{\frac{3(1+\alpha)}{n}}}{c-a^{\frac{3(1+\alpha)}{n}}}-\alpha}}
dx
\end{equation}
\subsection{Polytropic gas K-essence model}
The idea of the K-essence scalar field was motivated from the
Born-Infeld action of string theory  and can explain the late time
acceleration of the universe \cite{c26}. The general scalar field
action for K-essence
model as a function of $\phi $ and $\chi =\dot{\phi}^{2}/2$ is given by \cite%
{c27}
\begin{equation}
S=\int d^{4}x\sqrt{-g}\text{ }p(\phi ,\chi ),
\end{equation}%
where the Lagrangian density $p(\phi ,\chi )$ relates to a pressure
density and energy density through the following equations:
\begin{equation}
p(\phi ,\chi )=f(\phi )(-\chi +\chi ^{2}),
\end{equation}%
\begin{equation}
\rho (\phi ,\chi )=f(\phi )(-\chi +3\chi ^{2}).
\end{equation}%
Hence, the EoS parameter of K-essence scalar field is obtained as
\begin{equation}
\omega _{K}=\frac{p(\phi ,\chi )}{\rho (\phi ,\chi )}=\frac{\chi -1}{3\chi -1%
}.  \label{w_k}
\end{equation}%
By comparing Eqs.(\ref{eos1}) and (\ref{w_k}), we have
\begin{equation}
w_{\Lambda}=-1-\frac{a^{\frac{3(1+\alpha)}{n}}}{c-a^{\frac{3(1+\alpha)}{n}}}-\alpha=\frac{\chi-1}{3\chi-1}
\end{equation}
Hence the parameter $\chi$ is obtained as
\begin{equation}
\chi=\frac{2+\frac{a^{\frac{3(1+\alpha)}{n}}}{c-a^{\frac{3(1+\alpha)}{n}}}+\alpha}
{4+3\frac{a^{\frac{3(1+\alpha)}{n}}}{c-a^{\frac{3(1+\alpha)}{n}}}+3\alpha}
\end{equation}
From Eq.(\ref{w_k}), one can see the phantom behavior of K-essence scalar field ($w_{K}<-1$) when the parameter $\chi$ lies in the interval
$1/3<\chi<1/2$.

Using $\dot{\phi}^2=2\chi$ and changing the time derivative to the
derivative with respect to $x=\ln{a}$, we obtain
\begin{equation}\label{primephi}
\phi^{\prime}=\frac{1}{H}\sqrt{\frac{4+2\frac{a^{\frac{3(1+\alpha)}{n}}}{c-a^{\frac{3(1+\alpha)}{n}}}+2\alpha}
{4+3\frac{a^{\frac{3(1+\alpha)}{n}}}{c-a^{\frac{3(1+\alpha)}{n}}}+3\alpha}}
\end{equation}
The integration of Eq.(\ref{primephi}) yields
\begin{equation}
\phi(x)-\phi(0)=\int^{x}_{0}\frac{1}{H}\sqrt{\frac{4+2\frac{a^{\frac{3(1+\alpha)}{n}}}{c-a^{\frac{3(1+\alpha)}{n}}}+2\alpha}
{4+3\frac{a^{\frac{3(1+\alpha)}{n}}}{c-a^{\frac{3(1+\alpha)}{n}}}+3\alpha}}dx
\end{equation}
Here, we reconstructed the potential and the dynamics of K-essence
scalar field according to the evolutionary form of the interacting
polytropic gas model. The K-essence polytropic gas model can explain
the accelerating universe and also behaves as a phantom model
provided $1/3<\chi<1/2$.
\subsection{Polytropic gas dilaton model}
A dilaton scalar field can also be assumed as a source of DE. This
scalar field is originated from the lower-energy limit of string
theory \cite{c28}. The dilaton filed is described by the effective
Lagrangian density as
\begin{equation}
p_{D}=-\chi +ce^{\lambda \phi }\chi ^{2},
\end{equation}%
where $c$ and $\lambda $ are positive constant. Considering the
dilaton field as a source of the energy-momentum tensor in Einstein
equations, one can find that the Lagrangian density corresponds to
the pressure of the scalar field and the energy density of dilaton
field is also obtained as
\begin{equation}
\rho _{D}=-\chi +3ce^{\lambda \phi }\chi ^{2},  \label{rhod}
\end{equation}%
Here $2\chi =\dot{\phi}^{2}$. The negative coefficient of the kinematic term
of the dilaton field in Einstein frame makes a phantom like behavior for
dilaton field. The EoS parameter of dilaton is given by
\begin{equation}
\omega _{D}=\frac{p_{D}}{\rho _{D}}=\frac{-1+ce^{\lambda \phi }\chi }{%
-1+3ce^{\lambda \phi }\chi }.  \label{dilaton2}
\end{equation}%
In order to consider the dilaton field as a description of
polytropic gas, we establish the correspondence between the dilaton
EoS parameter,$w_{D}$, and the EoS parameter $w_{\Lambda }$ of
polytropic gas model. By equating Eq.(\ref{dilaton2}) with
Eq.(\ref{eos1}), we find
\begin{eqnarray}\label{dil1}
&&ce^{\lambda \phi }\chi=\frac{w_{\Lambda }-1}{3w_{\Lambda }-1}
=\frac{2+\frac{a^{\frac{3(1+\alpha)}{n}}}{c-a^{\frac{3(1+\alpha)}{n}}}+\alpha}
{4+3\frac{a^{\frac{3(1+\alpha)}{n}}}{c-a^{\frac{3(1+\alpha)}{n}}}+3\alpha}
\end{eqnarray}%
By using $\chi =\dot{\phi}^{2}/2$ and $\dot{\phi}=\phi ^{\prime }H$,
the scalar field $\phi$ can be obtained as
\begin{eqnarray}\label{dil3}
&&\phi (x) =\frac{2}{\lambda }\ln \Big(e^{\lambda \phi
(0)/2}+\frac{\lambda }{\sqrt{2c}}\int_{0}^{x}\frac{1}{H}
\sqrt{\frac{2+\frac{a^{\frac{3(1+\alpha)}{n}}}{c-a^{\frac{3(1+\alpha)}{n}}}+\alpha}
{4+3\frac{a^{\frac{3(1+\alpha)}{n}}}{c-a^{\frac{3(1+\alpha)}{n}}}+3\alpha}}dx\Big)
\end{eqnarray}%
Here we presented the reconstructed potential and dynamics of
dilaton scalar field according to the evolution of interacting
polytropic gas model.
\section{Conclusion}
In this work we presented the interacting polytropic gas model of
dark energy to interpret the accelerated expansion of the universe.
Assuming a non-flat FRW universe dominated by interacting polytropic
gas DE and CDM, we studied the cosmic behavior of polytropic gas
model. For this aim, we calculated the evolution of effective EoS
parameter and showed that for positive values $c>a^{3(1+\alpha)/n}$
with even numbers of $n$ this model behaves as a phantom DE model
and in the case of $-\infty<c\leq -a^{3/n}/2$ it treats as a
quintessence model. Similar to interacting agegraphic dark energy
model, the interacting polytropic gas model crosses the phantom
divide from below ($w_{\Lambda}<-1$) to up ($w_{\Lambda}>-1$). The
transition from phantom to quintessence depends on the parameter $c$
of the model. For larger value of $c$, the transition take place
sooner. In the case of phantom polytropic gas model, $w_{\Lambda}$
is larger for larger value of $c$. In the scenario of polytropic gas
model, the phantom divide can be crossed even in the absence of
interaction. We also calculated the evolution of energy density
$\Omega_{\Lambda}$. The matter dominated phase at the early time and
DE-dominated universe at the late time can be described in the
context of polytropic gas model. The polytropic gas starts to be
effective earlier for larger value of interaction parameter as well
as for larger value of the parameter $c$ or smaller value of $n$. We
calculated the deceleration parameter $q$ and obtained the
decelerated and accelerated expansion phases of the universe in the
context of polytropic gas model. The transition from decelerated
expansion ($q>0$) to accelerated expansion ($q<0$) takes place
sooner for larger value of $c$ and also by increasing the
interaction parameter $\alpha$. Since the scalar fields models are
the underlying theory of dark energy, we proposed a correspondence
between interacting polytropic gas model with the tachyon, K-essence
and dilaton scalar fields models. We reconstructed the potential and
the dynamics of these scalar fields according to the evolution of
the interacting
polytropic gas model.\\

\acknowledgments{This work has been supported by Research Institute
for Astronomy and Astrophysics of Maragha, Iran.}

\newpage
\begin{center}
\begin{figure}[!htb]
\includegraphics[width=8cm]{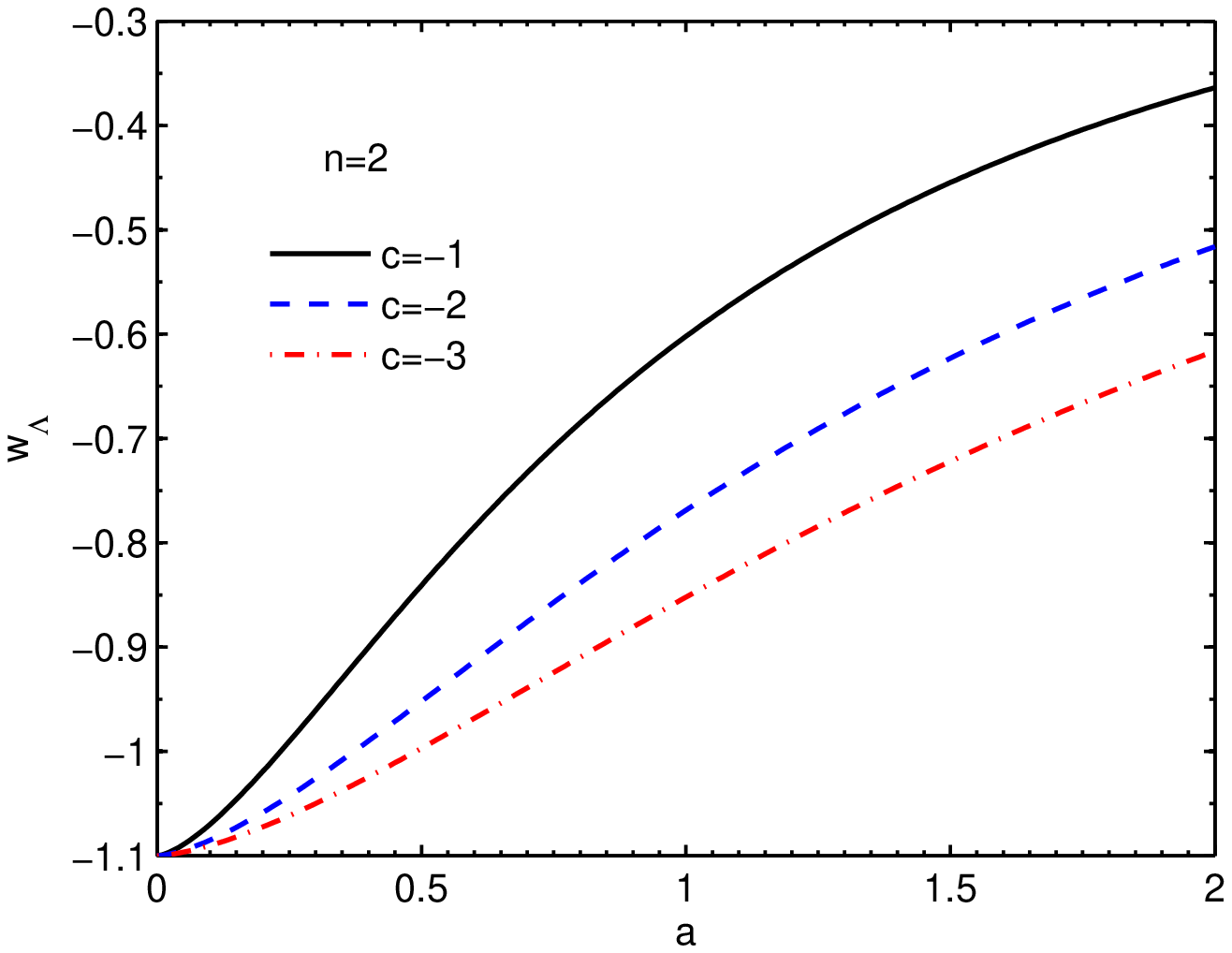} \includegraphics[width=8cm]{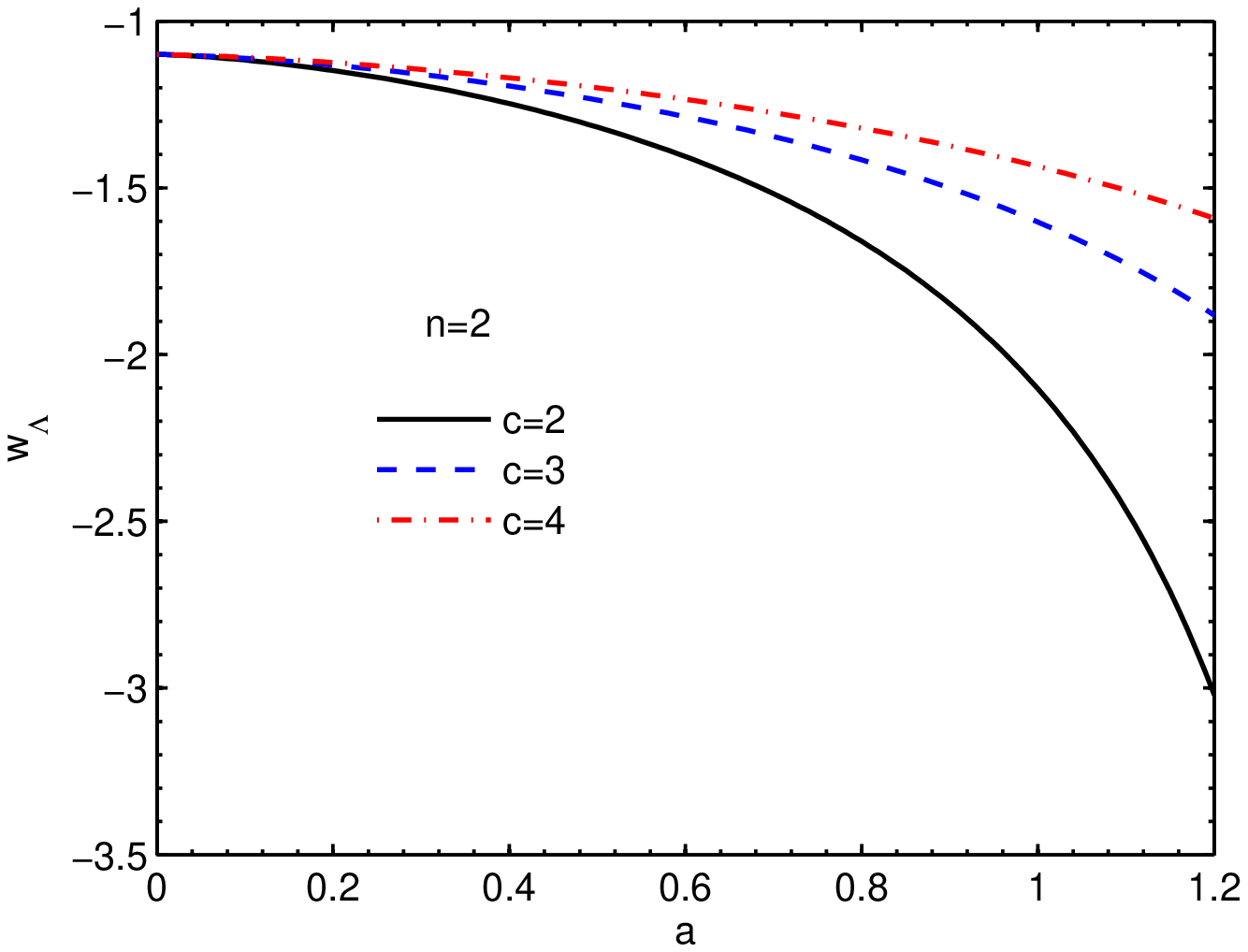}
\includegraphics[width=8cm]{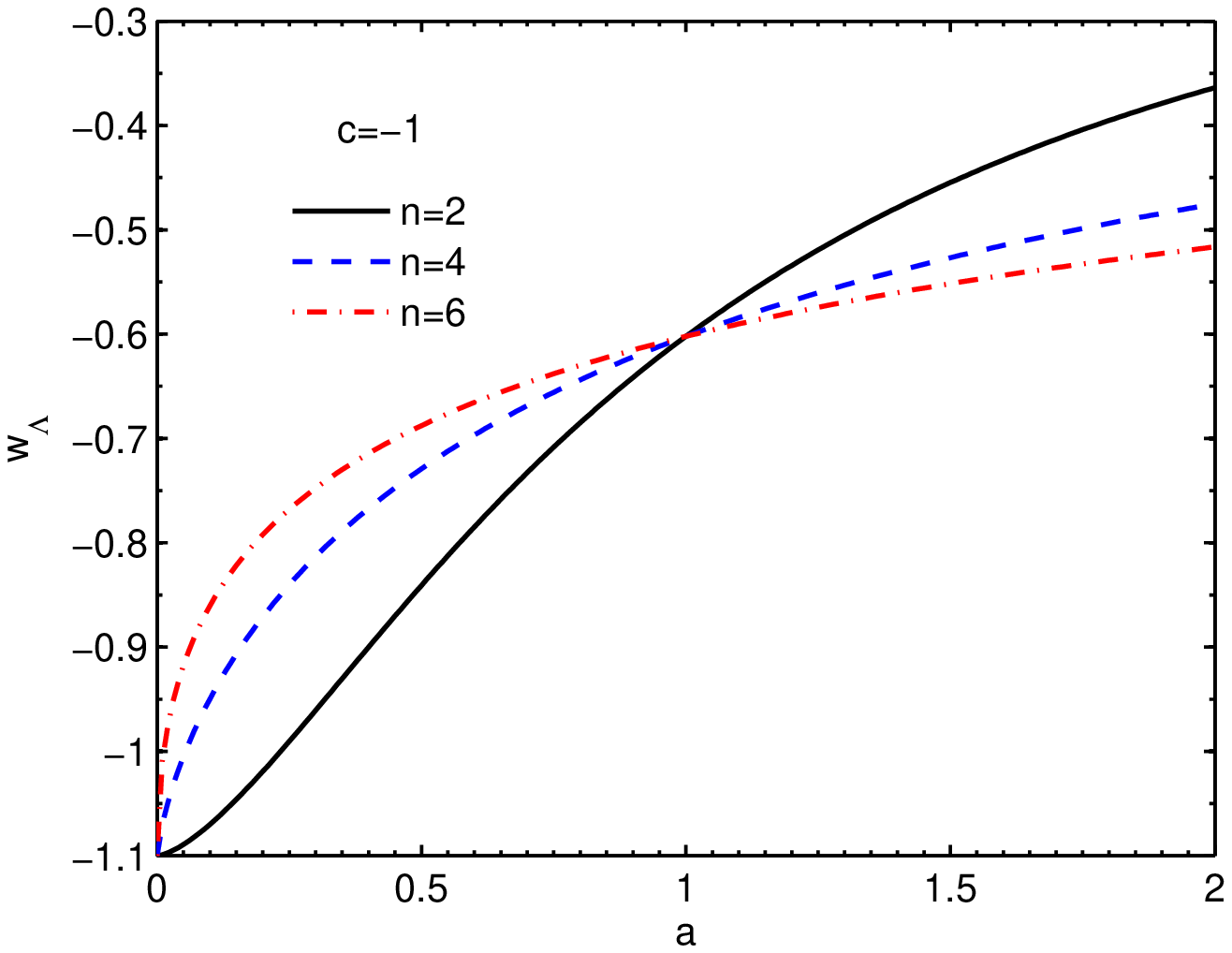}  \includegraphics[width=8cm]{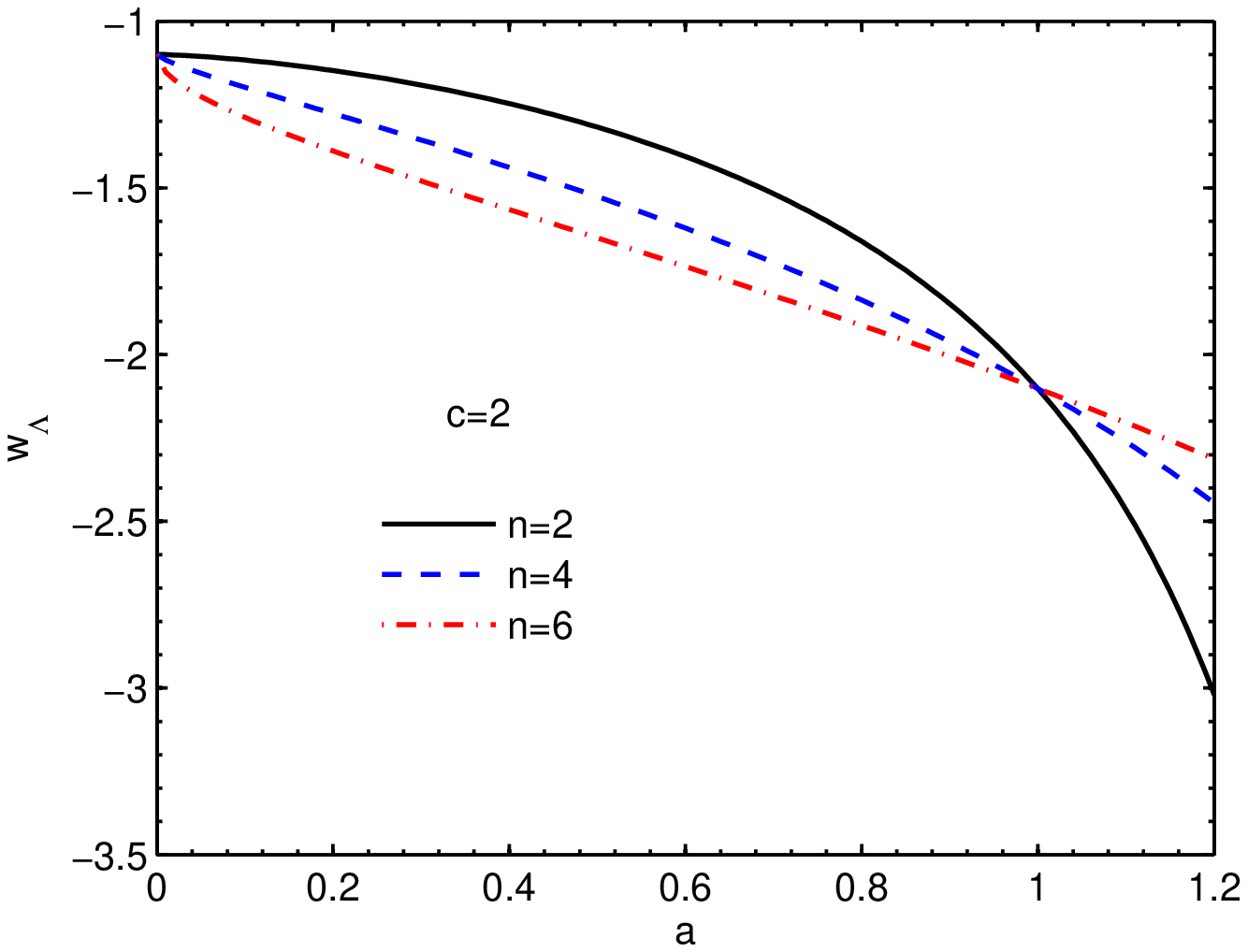}%
\caption{The  EoS parameter $w_{\Lambda}$ of interacting polytropic
gas model as a function of cosmic scale
factor $a$ for different model parameters $c$ and $n$. The interaction parameter is chosen as $\alpha=0.1$. \\[0pt]}
\end{figure}
\end{center}

\newpage

\begin{center}
\begin{figure}[!htb]
\includegraphics[width=8cm]{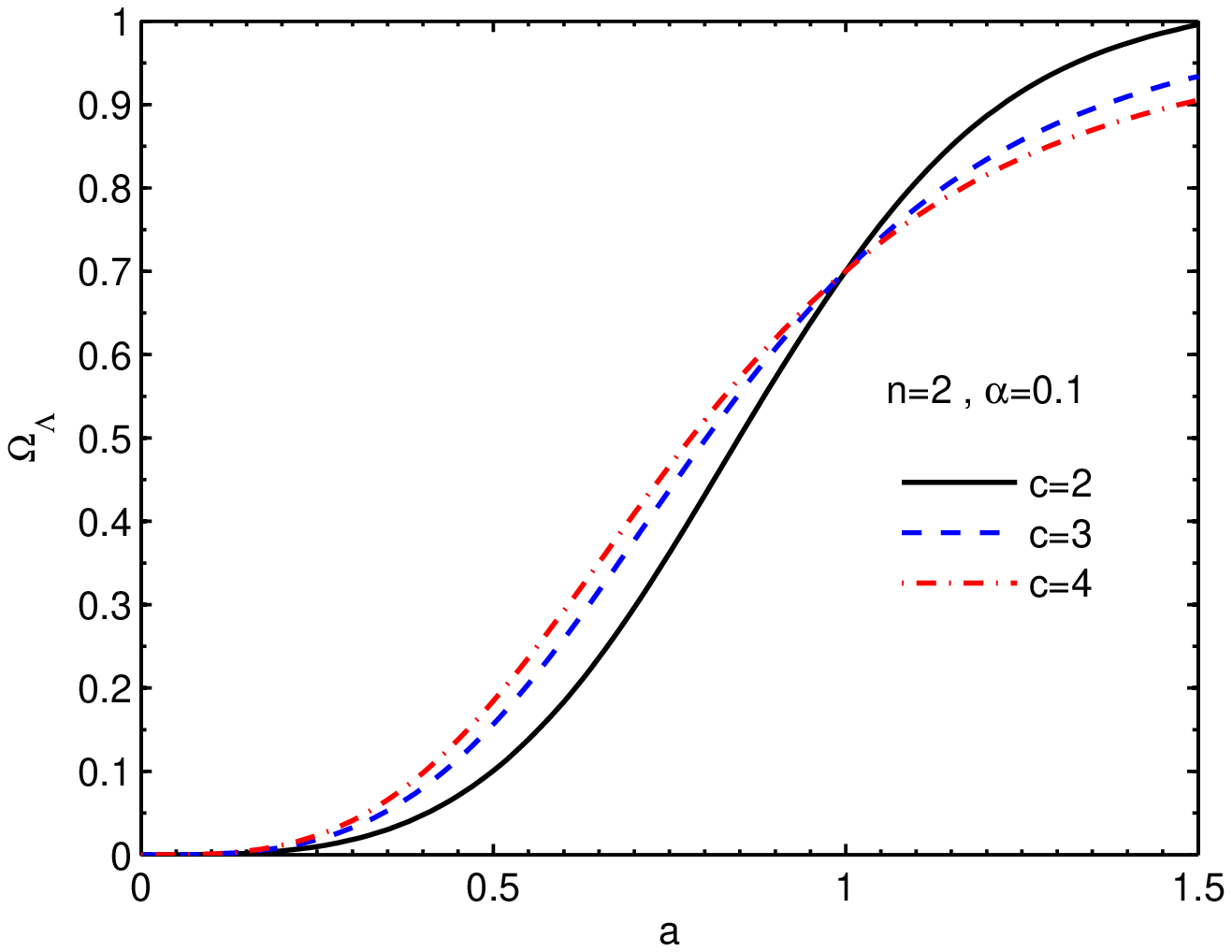} \includegraphics[width=8cm]{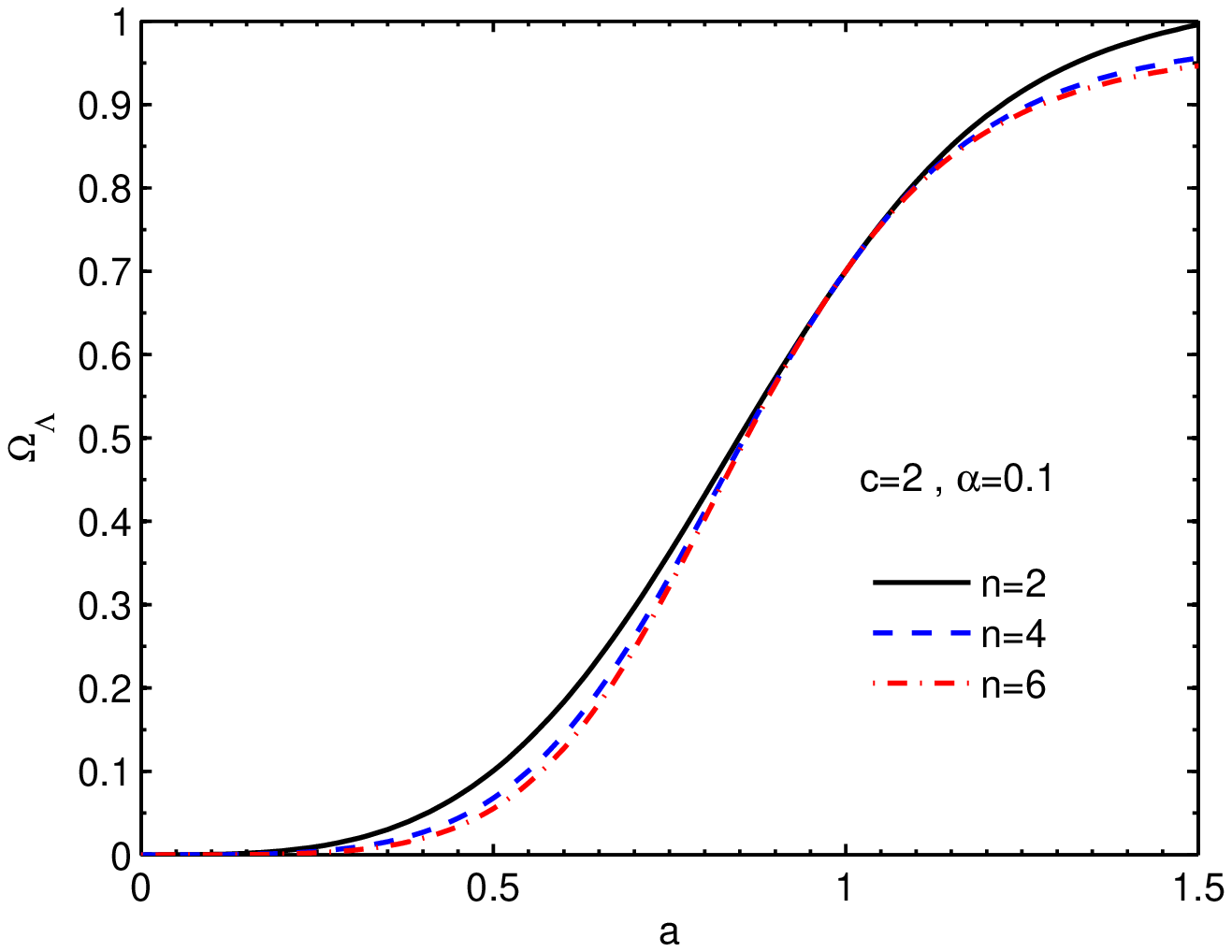}
\includegraphics[width=8cm]{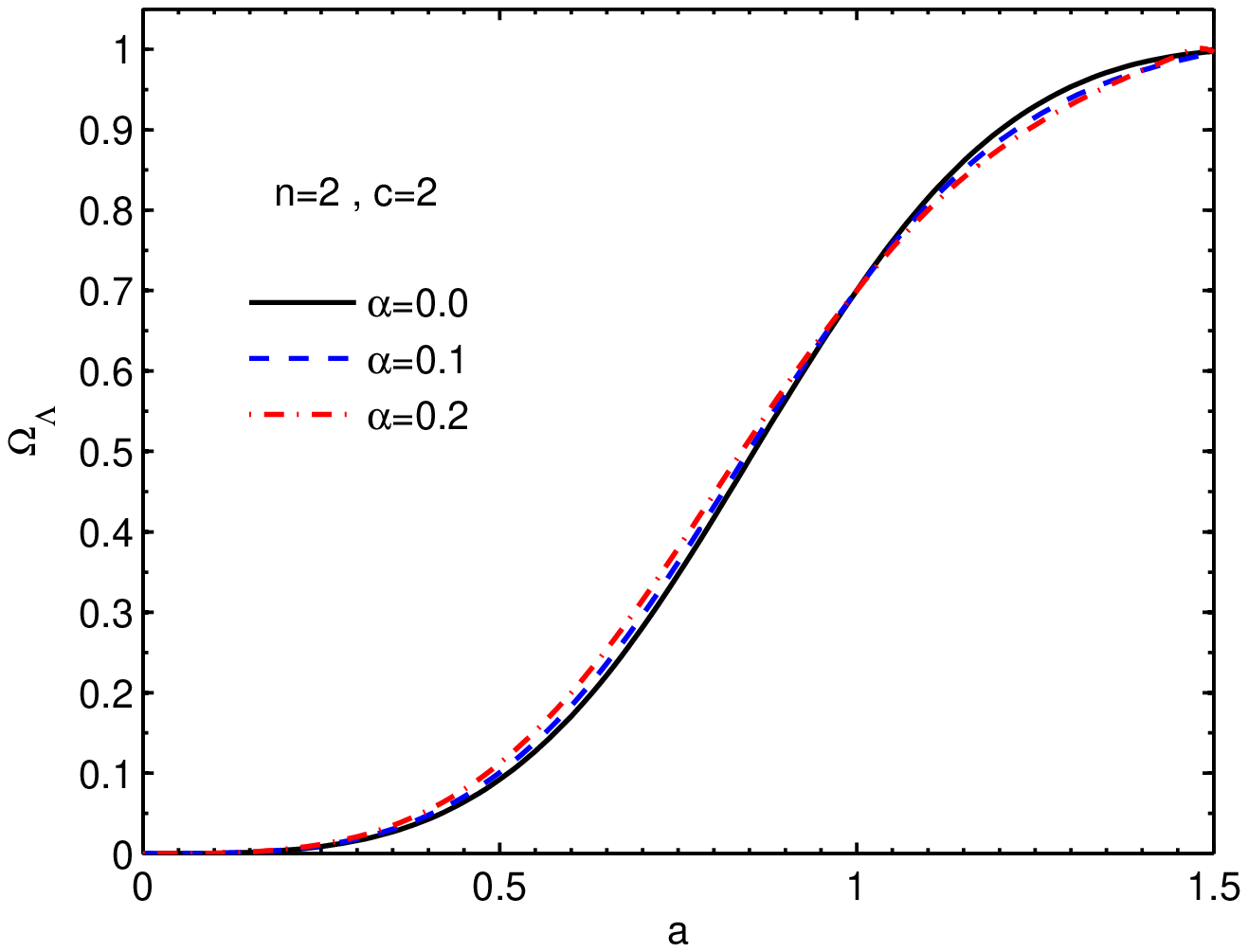} %
\caption{The evolution of energy density ,$\Omega_{\Lambda}$, in
terms of cosmic scale factor $a$ for interacting polytropic gas
model. In upper left panel, the parameter $n$ is fixed and the
parameter $c$ is varied. In upper right panel, we fix $c$ and vary
$n$. In lower panel, by fixing the parameters $c$
and $n$, we vary the interaction parameter $\alpha$.\\[0pt]}
\end{figure}
\end{center}

\begin{center}
\begin{figure}[!htb]
\includegraphics[width=8cm]{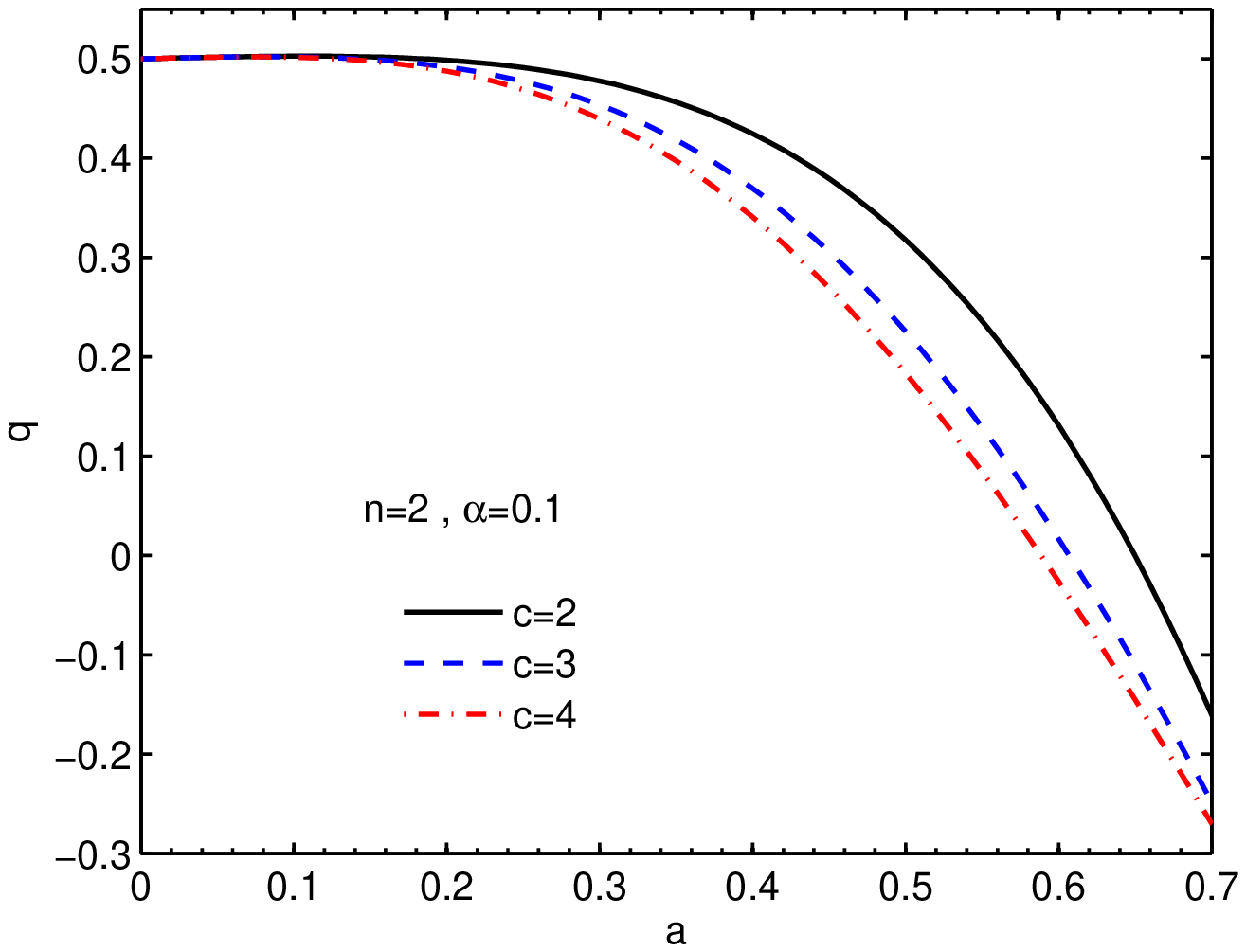} \includegraphics[width=8cm]{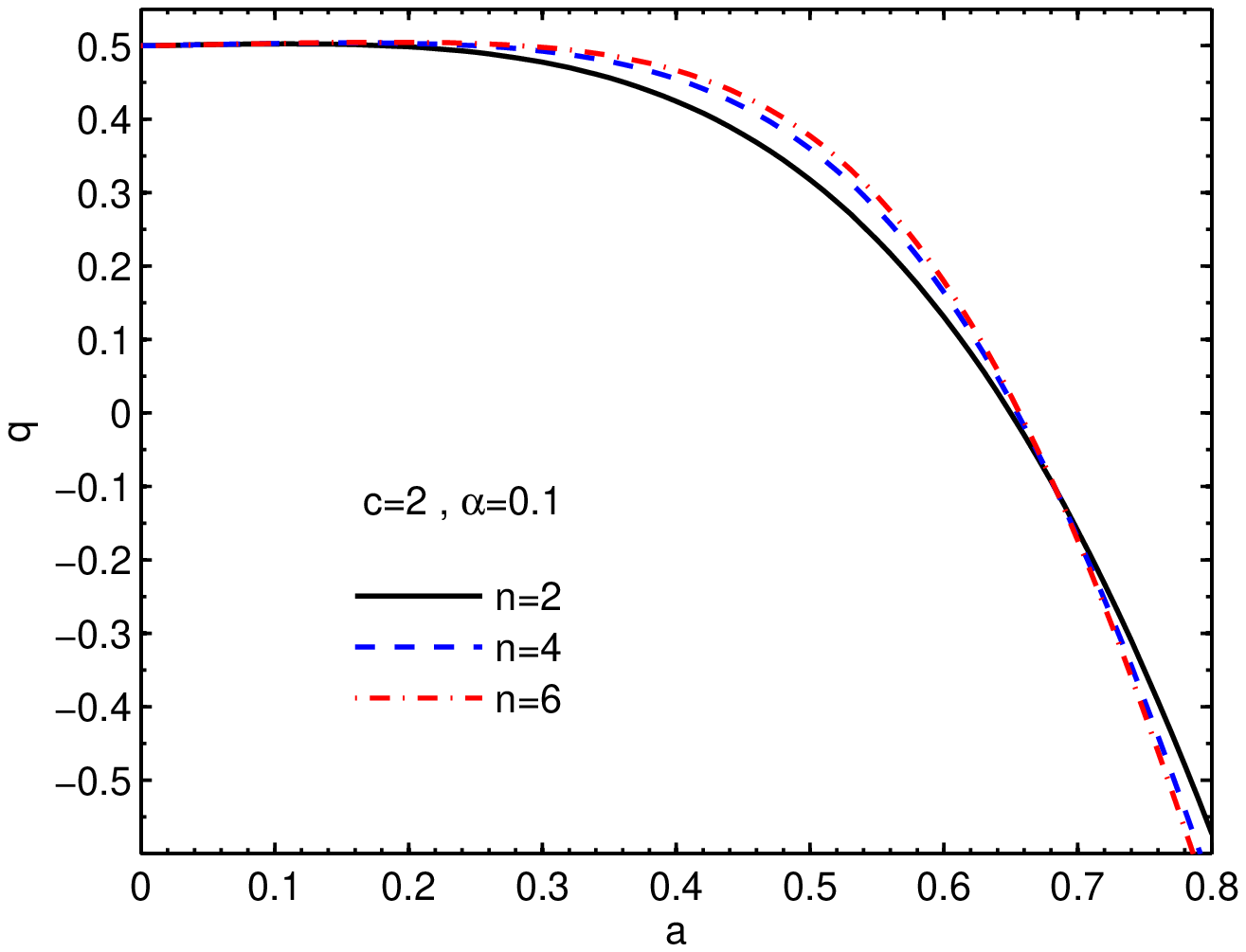}
\includegraphics[width=8cm]{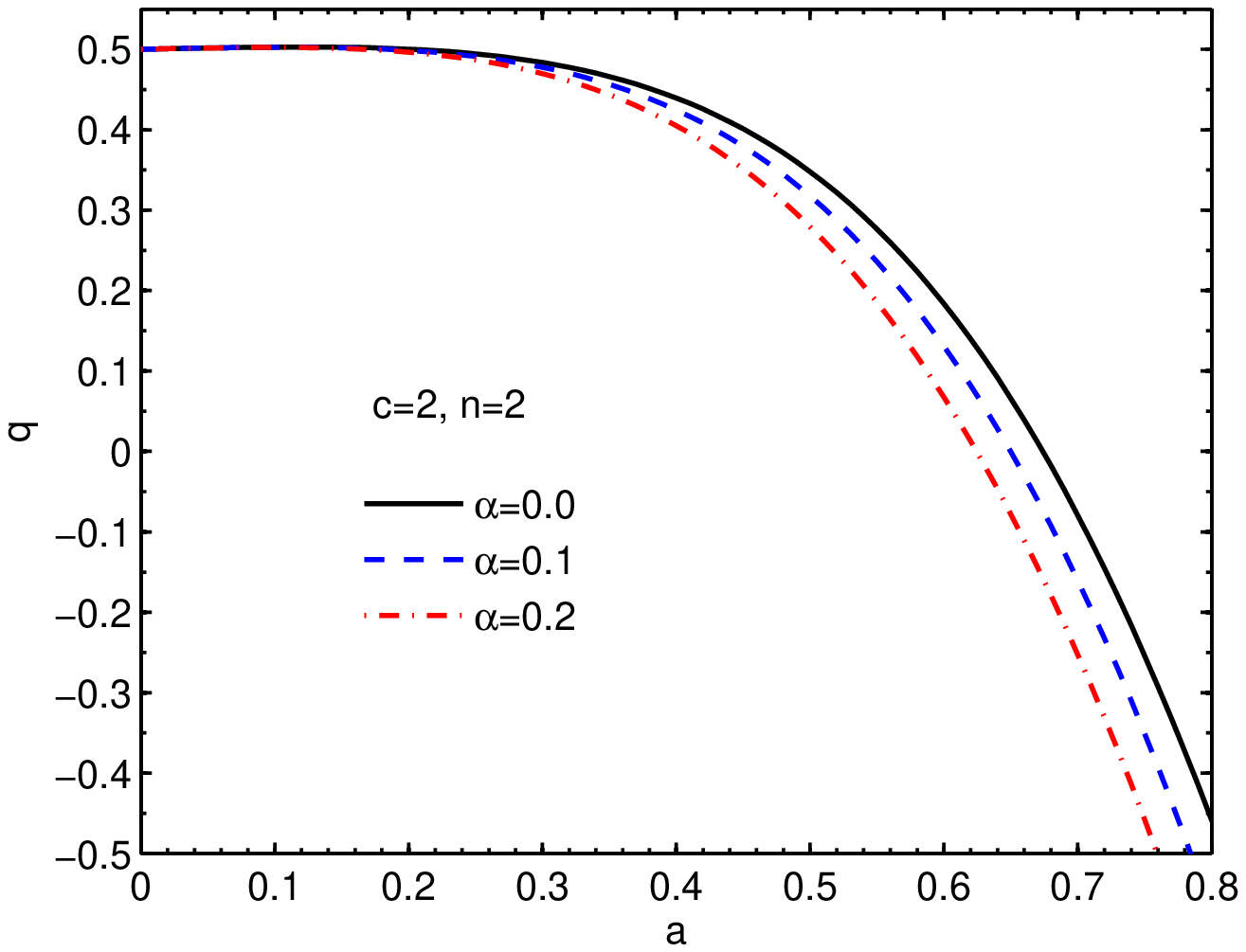} %
\caption{The evolution of deceleration parameter $q$ versus of scale
factor $a$ for interacting polytropic gas model. In upper left
panel, the parameter $n$ is fixed and the parameter $c$ is varied.
In upper right panel, we fix $c$ and vary $n$. In lower panel, by
fixing the parameters $c$ and $n$, we vary
the interaction parameter $\alpha$.\\[0pt]}
\end{figure}
\end{center}

\end{document}